\documentclass[aps,prl,amsmath,showpacs,twocolumn,amssymb,footinbib,superscriptaddress]{revtex4}

\usepackage{amssymb,amsfonts,amsmath}

\usepackage[apple mac]{inputenc}
\usepackage[english]{babel}
\usepackage{times}
\usepackage{latexsym}
\usepackage{fancyhdr}
\usepackage{verbatim}
\usepackage{graphicx}
\usepackage{epsf}
\usepackage{subfigure}
\usepackage{bm}
\usepackage{epstopdf}

\def\be{\begin{equation}}
\def\ee{\end{equation}}
\def\ba{\begin{align}}
\def\enda{\end{align}}
\def\bi{\begin{itemize}}
\def\ei{\end{itemize}}
\def\bs#1{\boldsymbol{#1}}
\def\txt#1{\textrm{#1}}

\def\nm{{\ {\rm nm}}}						
\def\micron{{\ \mu{\rm m}}}					

\def\Hz{{\ {\rm Hz}}}						
\def\kHz{{\ {\rm kHz}}}						
\def\MHz{{\ {\rm MHz}}}						

\def\nK{{\ {\rm nK}}}						

\def\Rb87{^{87}\rm{Rb}}						
\def\Li6{^{6}\rm{Li}}						

\newcommand{\ket}[1]{|#1\rangle}

\begin{document}

\title{Realistic Time-Reversal Invariant Topological Insulators With Neutral Atoms}

\author{N.~Goldman}
\affiliation{Center for Nonlinear Phenomena and Complex Systems - Universit$\acute{e}$ Libre de Bruxelles (U.L.B.), Code Postal 231, B-1050 Brussels, Belgium}

\author{I.~Satija}
\affiliation{
Department of Physics, George Mason University, Fairfax, Virginia 22030, USA}
\affiliation{National Institute of Standards and Technology (NIST), Gaithersburg, Maryland 20899, USA}

\author{P.~Nikolic}
\affiliation{
Department of Physics, George Mason University, Fairfax, Virginia 22030, USA}
\affiliation{National Institute of Standards and Technology (NIST), Gaithersburg, Maryland 20899, USA}

\author{A.~Bermudez}
\affiliation{
Departamento de F\'isica Te\'orica I, Universidad Complutense, 28040 Madrid, Spain}

\author{M.~A.~Martin-Delgado}
\affiliation{
Departamento de F\'isica Te\'orica I, Universidad Complutense, 28040 Madrid, Spain}

\author{M.~Lewenstein}
\affiliation{
ICFO-Institut de Ci\`encies Fot\`oniques, Parc Mediterrani de la Tecnologia, E-08860 Castelldefels (Barcelona), Spain}
\affiliation{ICREA - Instituciio Catalana de Recerca i Estudis Avancats, 08010 Barcelona, Spain}

\author{I.~B.~Spielman}
\affiliation{Joint Quantum Institute, NIST and University of Maryland, Gaithersburg, Maryland, 20899, USA}

\begin{abstract} 
We lay out an experiment to realize time-reversal invariant topological insulators in alkali atomic gases. We introduce an original method to synthesize a gauge field in the near-field of an atom-chip, which effectively mimics the effects of spin-orbit coupling and produces quantum spin-Hall states. We also propose a feasible scheme to engineer sharp boundaries where the hallmark edge states  are localized.  Our multi-band system  has a large parameter space exhibiting a variety of quantum phase transitions between topological and normal insulating phases. Due to their remarkable versatility, cold-atom systems are ideally suited to realize topological states of matter and drive the development of topological quantum computing.


\end{abstract}

\pacs{67.85.-d, 81.16.Ta,37.10.Jk}

\maketitle

Topological insulators are a broad class of unconventional materials that are insulating in the interior but conduct along the edges. The edge transport  is topologically protected and dissipationless.  This subject has emerged as a new frontier, discovering novelties in the single-particle band theory and
providing a new impetus to the many-body physics of strongly correlated systems. Until recently, the only known topological insulators -- quantum Hall (QH) states -- violated time-reversal (TR) symmetry.  However, the discovery of the quantum spin Hall (QSH) effect demonstrated the  existence of novel topological states not rooted in time-reversal violations~\cite{Kane2005bis,Bernevig2006bis,edgebis}, and has opened the possibility to design new spintronic devices exploiting the spin-dependent currents carried by the edge states.

Realizing topological insulators with cold atoms is particularly attractive, and setups exploring the TR-breaking case have been envisaged~\cite{Wu2008,Stanescuarxiv}. In this Letter, we propose a concrete setup using fermionic $^{6}$Li subjected to a synthetic gauge field, which provides an archetypical system for investigating the QSH phase. There are numerous proposals for engineering gauge fields~\cite{Jaksch2003}, which generally depend on laser-induced Raman coupling between internal atomic states. Such a method was recently implemented~\cite{Lin2009bis} for bosonic $^{87}$Rb atoms, but would lead to large spontaneous emission rates for the alkali fermions.  Here we describe a setup that combines state-independent optical potentials with micron-scale state-dependent magnetic potentials in an atom chip. By completely eliminating spontaneous emission, this approach makes practical the realization of gauge fields for all alkali atoms. We demonstrate that these synthetic fields lead to the purest realization of the QSH effect and allow to explore striking aspects of this topological state of matter. In particular, the stability of the topological phases against interactions can be explored by means of Feschbach resonances.

In condensed-matter systems, the  QSH effect originates from a material's intrinsic spin-orbit coupling~\cite{Kane2005bis,Bernevig2006bis}. Note that such a coupling is analogous to a non-Abelian gauge field $\bs{\mathcal{A}}={\bf A} \check\sigma_{\bf z}$, where $\check\sigma_{\bf{x},\bf{y},\bf{z}}$ are the Pauli matrices. This observation emphasizes that the QSH effect consists of spin-1/2 fermions where the two spin components are described as QH states at equal but opposite ``magnetic fields.'' We hereby demonstrate how to synthesize such a gauge field in an optical lattice and show how it leads to QSH physics. 

\begin{figure}[tbp]
\begin{center}
\includegraphics[width=3.3in]{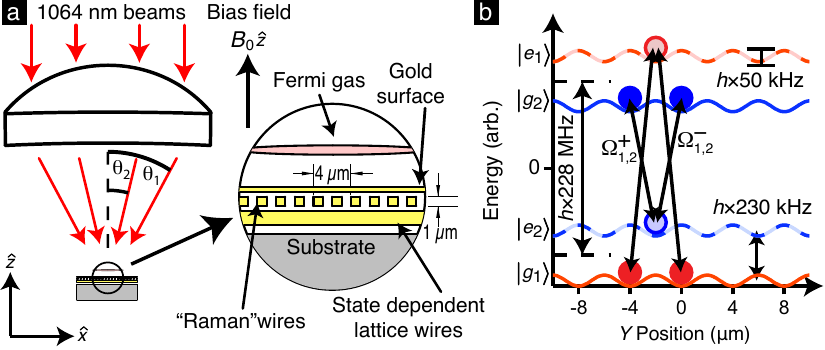}
\end{center}
\caption{ {\bf (a)} Experimental layout showing the origin of optical (state independent) and magnetic (state dependent) potentials and coupling fields.  A state-independent, staggered, lattice along $\hat x$ is formed by the separate interference of two pairs of $\lambda\!=\!1064\nm$ laser beams slightly detuned from each other to eliminate cross interference.  The respective intersection angles are chosen so the lattice period differ by a factor of two.  Both beams reflect from the chip-surface and form vertically aligned lattices, trapping the degenerate Fermi gas about $5\micron$ above the surface.  The inset shows the chip geometry, from top to bottom: a reflective chip-surface, gold wires aligned along $\hat y$ placed every $2\micron$ along $\hat x$ (producing the $\ket{g}$-$\ket{e}$ coupling), and finally gold wires aligned along $\hat x$ with a $2\micron$ spacing (producing the state dependent lattice).  {\bf (b)} Computed atomic potentials and radio-frequency (RF) driven ``Raman'' transitions.}
\label{FIG1}
\end{figure}

Our proposal for realizing a fermionic model with a SU(2) gauge structure requires four atomic states $\ket{g_1} = \ket{F\!=\!1/2,m_F\!=\!1/2}$, $\ket{g_2} = \ket{3/2,-1/2}$, $\ket{e_1} = \ket{3/2,1/2}$, and $\ket{e_2} = \ket{1/2,-1/2}$, in a square lattice described by the Hamiltonian
\begin{align}
\mathcal{H}=&-t\sum_{m,n} \bs c_{m+1,n}^{\dagger} e^{i \check\theta_{{\bf x}}} \bs c_{m,n}
+\bs c_{m,n+1}^{\dagger} e^{i \check\theta_{\bf y}} \bs c_{m,n}+\txt{h.c.}  \notag \\
&+  \lambda_{\txt{stag}} \,  \sum_{m,n}  (-1)^{m} \, \bs c^{\dagger}_{m,n} \bs c_{m,n}.
\label {2dh}
\end{align}
$\bs c_{m,n}$ is a 2-component field operator defined on a lattice site $(x\!=\!ma,\ y\!=\!n a)$, $a$ is the lattice spacing, $m$, $n$ are integers, and $t$ is the nearest-neighbor hopping.  In this model, the Peierls phases $\check\theta_{{\bf x},{\bf y}}$ result from a synthetic gauge field~\cite{Jaksch2003} that modifies the hopping along $\hat x$ and $\hat y$. Here, all the states experience a primary lattice potential $V_1(x)=V_x\sin^2(k x)$ along $\hat x$ which gives rise to a hopping amplitude $t\!\approx\!0.4\kHz$.  A secondary much weaker lattice $V_2(x)=2\lambda_{\txt{stag}}\sin^2(k x/2)$ slightly staggers the primary lattice with $\lambda_{\txt{stag}} \approx t$.  These lattices, with an approximate period of $a\!=\!2\micron$, are produced by two pairs of $\lambda\!=\!1064\nm$ lasers, slightly detuned from each other and incident on the atom chip's reflective surface [Fig. ~\ref{FIG1}(a)]. Additionally, these beams create a lattice along $\hat z$ with a $0.55\micron$ period, confining the fermions to a 2D plane.

We study the SU(2) hopping operators
\begin{align}
\check\theta_{\bf x} &= 2\pi \gamma {\bf \check\sigma_x},&& &\check\theta_{\bf y} &=  2\pi x \alpha {\bf \check\sigma_z},
\label{gaugefields}
\end{align}
where we set $a\!=\!\hbar\!=\!1$. The hopping operator $\check\theta_{\bf y} $ corresponds to opposite ``magnetic fluxes'' $ \pm \alpha$ for each spin component, whereas $\check\theta_{\bf x} $ mixes the spins. Our setup thus provides a SU(2) generalization of the well-known two-dimensional electron gas in a magnetic field~\cite{Hofstadter1976}. In order to engineer these state-dependent tunnelings, the states $\ket{g}$ and $\ket{e}$ must experience oppositely-signed lattices along $\hat y$ [Fig.~\ref{FIG1}(b)]. This can be directly implemented with the Zeeman shift $g\mu_B \left|B\right|$ of atoms provided that $\ket{g_{1,2}}$ and $\ket{e_{1,2}}$ have equal, but opposite magnetic moments $g$.  In our proposal, the magnetic moments are correctly signed and differ by less than 1\% in magnitude at a bias field $B\!=\!0.25\ {\rm G}$. With these states, a state-dependent lattice potential can be generated by an array of current-carrying wires with alternating $+I$ and $-I$ currents, spaced by a distance $a$.  A modest $I\!=\!5\ \mu{\rm A}$ current~\cite{Trinker2008} in wires $3\micron$ below the chip-surface produces a $6\ E_L$ Zeeman lattice [$E_L = h^2/8Ma$ is the lattice recoil energy], with a negligible $3\Hz$ hopping matrix element. The assisted-hopping along $\hat y$, with an $x$-dependent phase, can be realized with an additional grid of wires spaced by $a\!=\!2\micron$ along $\hat x$, with currents $I_m$. This provides {\it moving} Zeeman lattices with wave-vector $q$, where $I_m = I_0 \sin(q m a  - \omega t)$, leading to effective ``Raman couplings''.  The $\omega/2\pi \!\approx\! 228\pm0.23\MHz$ transitions indicated with arrows in Fig.~1 (b), are independently controllable in phase, amplitude, and wave-vector by commanding concurrent running waves at the indicated resonant frequencies.  The minimum wavelength $d=2\pi/q$ of this moving lattice is Nyquist limited by $d > 2a$. In the frame rotating at the angular frequency $\omega$, and after making the rotating wave approximation the coupling terms have the desired form $t \exp(i q m a)$.  A potential gradient along $\hat y$ detunes this Raman coupling into resonance and is produced by simply shifting the center of the harmonic potential.   In the model Hamiltonian~\eqref{2dh}-\eqref{gaugefields}, the phase for hopping along $\hat y$ is $\alpha=q a/2\pi = a/d$ in terms of physical parameters.  Our scheme also requires a contribution to the hopping along $\hat x$ that mixes the $\ket{e}$ and $\ket{g}$.  This can be realized using a Zeeman lattice moving along $\hat x$, but tuned to drive transitions between $\ket{g_1}\rightarrow\ket{g_2}$ and $\ket{e_1}\rightarrow\ket{e_2}$. The corresponding control parameter $\gamma$, and the additional staggered potential which induces alternate on-site energies $\epsilon=\pm \lambda_{\txt{stag}}$ along the $x$-axis, are shown below to enrich the QSH physics in a novel manner. Note that the most sensitive parameter is the resonance condition required for the ``Raman couplings". Its stability relies on the absolute control of the state-dependent potentials provided by the RF magnetic fields. 

The engineered Hamiltonian  \eqref{2dh}-\eqref{gaugefields} satisfies  TR invariance,  since it commutes with the TR-operator defined as $\mathcal{T}= i {\bf \check\sigma_y} K$, where $K$ is the complex-conjugate operator.  This synthetic, yet robust, TR symmetry enables the realization of $Z_2$-topological insulators in cold-atom laboratories.  We stress that the two components of the field operators correspond in general to a pseudo-spin 1/2, but in our proposal refer specifically to spin components of $^6$Li in its electronic ground state.

\begin{figure}[htb]
\begin{center}
\includegraphics[width=3.3in]{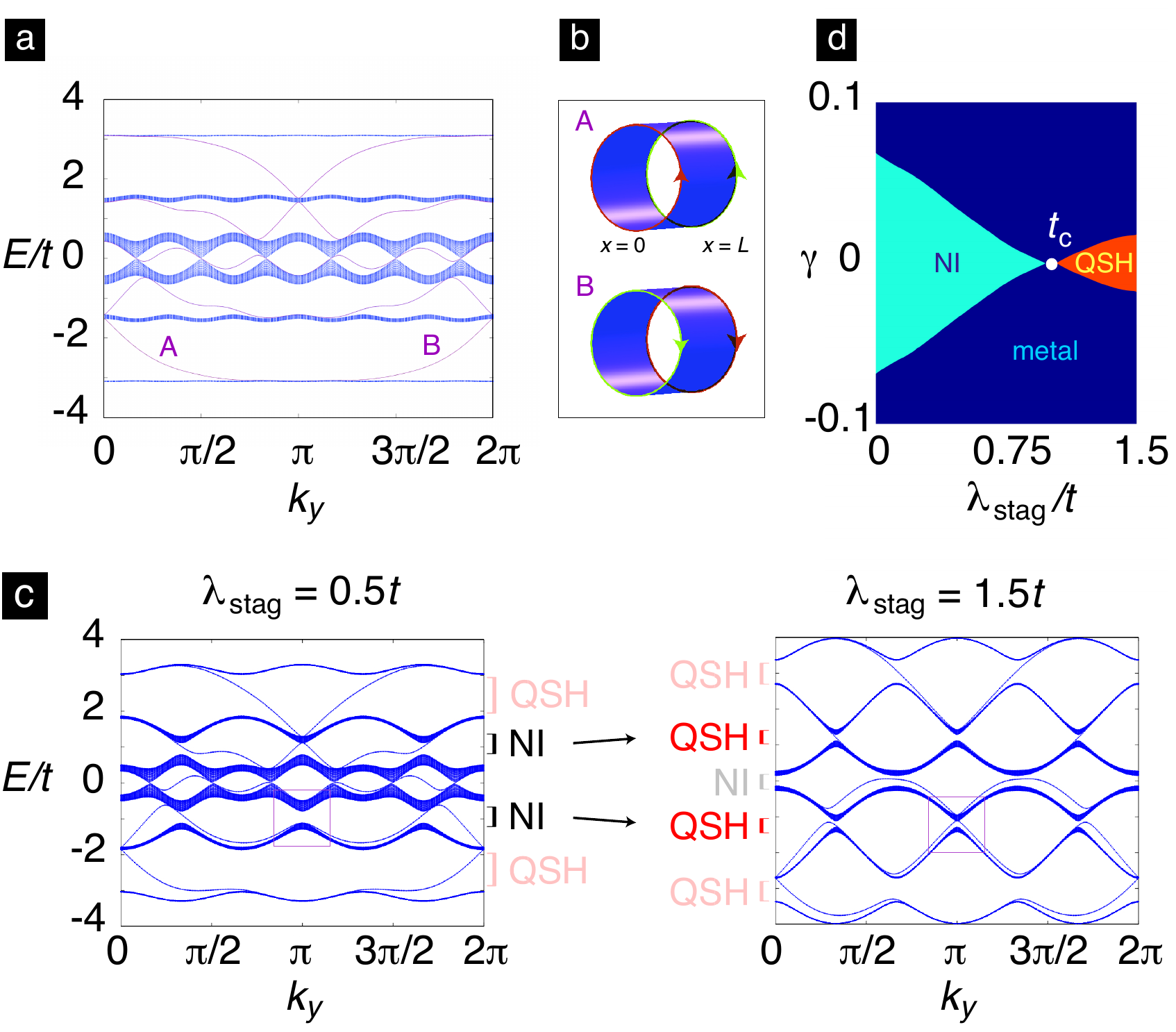}
\caption{\label{FIG2} {\bf (a)}  Energy spectrum of the uncoupled system ($\gamma=\lambda_{\txt{stag}}=0$ and $\alpha\!=\!1/6$) computed in a cylindrical geometry: the bulk energy bands (thick blue bands) are traversed by edge states (thin purple lines). {\bf (b)} Schematic representation of the two edge states, A and B, that lie inside the first bulk energy gap depicted in (a). The \emph{spins} traveling around the edges are respectively represented by red and  green arrows. {\bf (c)} Energy bands $E(k_y)$ for $\gamma\!=\!0$ and $\alpha\!=\!1/6$, with an external staggered potential $\lambda_{\txt{stag}}=0.5 t$ and $1.5 t$. The topological phases associated with the open bulk gaps are indicated. The purple rectangles highlight the NI to QSH phase transition that occurs around $E_F=-t$.  {\bf (d)} Phase diagram in the $(\gamma , \lambda_{\txt{stag}})$-plane in the vicinity of the uncoupled case $\gamma\!=\!0$ for $E_F = - t$.}
\end{center}
\end{figure}

In the absence of the confining trap $V_{\txt{conf}}(x,y)$,  the system can be solved on an abstract cylinder. We first study the nature of topological insulators with this partially closed geometry and then show how they can be detected when the trap is applied to the realistic open geometry. When $\gamma\!=\!0$, Eqs.~\eqref{2dh}-\eqref{gaugefields} describe two uncoupled  QH systems and for generic $\alpha=p/q$, where $p,q$ are integers, the fermion band-structure splits into $q$ subbands~\cite{Hofstadter1976}.  Our setup thus provides a multi-gap system, where a variety of band-insulators can be reached by varying the atomic filling factor. As discussed below, some of these insulators are topologically non-trivial and feature gapless edge states. The latter are localized at the boundaries of the sample, and  correspond to gapless excitations.  When the Fermi energy $E_F$ lies inside a bulk gap, the presence of these states is responsible for the spin transport along the edges. In this Letter, we present results for the case $\alpha\!=\!1/6$, which exhibits extremely rich phase diagrams with almost all possible topological phase transitions.  The topological phases discussed below are robust against small variations $\delta \alpha\!\sim\!0.01$ and rely on the existence of bulk gaps, which are continuously deformed when $\alpha$ is varied~\cite{Hofstadter1976}. Other configurations of the gauge field could be experimentally designed and would lead to similar effects. In Fig.~\ref{FIG2}(a), illustrating the spectrum for $\gamma\!=\!0$, the bulk bands are clearly differentiated from the gapless edge states within the bulk gaps. In the lowest bulk gap, the edge state channel A, and its TR-conjugate B, correspond to localized excitations which travel in opposite directions [Fig.~\ref{FIG2}(b)].  Thus the boundaries are populated by a single Kramers pair of counter-propagating states, each corresponding to an opposite spin: this lowest bulk gap describes a topological QSH phase.  Conversely, the next gap located at $E\!\approx\!-1$ is traversed by an even number of Kramers pairs: it is hence topologically equivalent to a normal band insulator (NI) \cite{Kane2005bis}. 

An alternative approach to the above even-odd criteria relies on the computation of the $Z_2$-topological invariant $\nu$ that characterizes the bulk gaps~\cite{Kane2005bis,edgebis}: $\nu(\txt{QSH})\!=\!1$ and $\nu(\txt{NI})\!=\!0$. When $\gamma\!=\!0$, spin is conserved and the spin-conductivity is quantized as $\sigma_s = e\nu/2 \pi$~\cite{Kane2005bis,Bernevig2006bis}. We verified that the four gaps depicted in Fig.\ref{FIG2}(a) are indeed associated to the sequence $\nu= \{ 1 , 0 , 0 , 1 \}$.  Surprisingly, topological insulators with distinct orders can be realized in this multi-band scenario by simply varying the atomic filling factor.   

In our multi-band system, the lattice-potential distortions can drive direct transitions between normal and topological insulating states.  Figure~\ref{FIG2}(c) shows the bulk gaps and edge states for successive values of the experimentally controllable staggered potential $\lambda_{\txt{stag}}$. We demonstrate that $\lambda_{\txt{stag}}$ induces a quantum phase transition (QPT) from a NI to a QSH phase even in the uncoupled case $\gamma\!=\!0$. This transition occurs within the  bulk gaps around $E = \pm 1$ at the critical value $\lambda_{\txt{stag}}=t$. At half-filling, the existence of Dirac points resist the opening of the gap for small staggered potential and eventually lead to a NI phase for $\lambda_{\txt{stag}}>1.25 t$. The $Z_2$-index analysis provides an efficient tool to obtain the full phase diagram in the $(\gamma , \lambda_{\txt{stag}})$-plane. The phase diagram represented in Fig.~\ref{FIG2}(d)  has been obtained numerically by evaluating the index $\nu$ inside the gap at $E \approx - t$, for small spin-mixing $\gamma < 0.1$. We observe three distinct phases: metallic (blue), QSH (red), and  NI (cyan).  These three phases coexist at a tri-critical point situated at $\gamma\!=\!0$ and $\lambda_{\txt{stag}}=t$.  The QSH phase occurs for a wide range of $\gamma$, indicating the robustness of this topological phase under small spin-mixing perturbations.

\begin{figure}[htb]
\begin{center}
\includegraphics[width=3.3in]{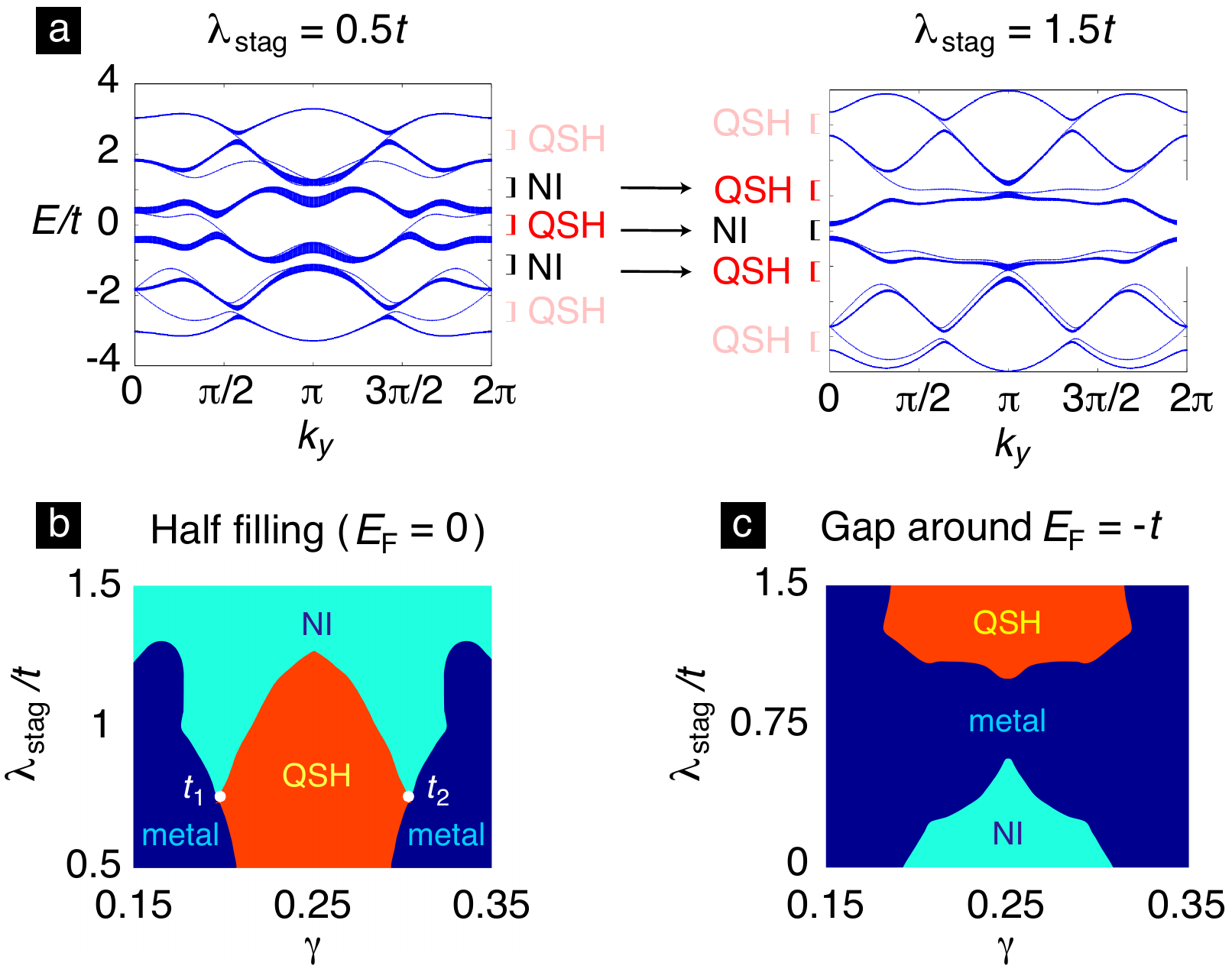}
\caption{\label{FIG3}  {\bf (a)} Energy bands $E(k_y)$ for $\gamma\!=\!0.25$ and $\alpha\!=\!1/6$, with an external staggered potential $\lambda_{\txt{stag}}=0.5 t$ and $1.5 t$. The topological phases associated to the bulk gaps are indicated.  {\bf (b)-(c)} Phase diagrams in the $(\gamma , \lambda_{\txt{stag}})$-plane in the vicinity of the maximally coupled case $\gamma\!=\!0.25$ for (b) $E_F\!=\!0$ and (c) $E_F\!=\!-t$.}
\end{center}
\end{figure}

The possibilities offered by cold-atom experiments enables us to consider the strong coupling regime corresponding to $\gamma\!=\!0.25$. In this limit, the previously independent QH subsystems ($\gamma\!=\!0$) become maximally coupled, drastically modifying the topological phase transitions presented above.  In Fig.~\ref{FIG3}(a), we illustrate the bulk bands and edge states for $\lambda_{\txt{stag}}=0.5 t$ and $1.5 t$, and the gaps are labeled according to the even-odd number of TR pairs. In this strong-coupled regime, a radically different scenario emerges: \emph{opposite} phase transitions occur successively in the neighboring gaps. First, gap-closings around $E \approx \pm 1$ occur and trigger NI to QSH phase transitions at $ \lambda_{\txt{stag}}= t$. Then, for $ \lambda_{\txt{stag}}=1.25 t$, the opposite transition QSH-NI occurs at half-filling ($E\!=\!0$). To fully capture the richness of this phenomenon, we numerically compute the index $\nu$ for a wide range of the parameters around $\gamma\!=\!0.25$.  At half-filling, the phase diagram features tri-critical points, and the QSH-NI phase transitions occur along a well-defined curve [Fig.~\ref{FIG3}(b)]. On the other hand, in the neighboring gaps, the NI to QSH phase transition is separated by an intermediate metallic region [Fig.~\ref{FIG3}(c)]. Therefore, by manipulating $\lambda_{\txt{stag}}$, $E_F$, and the coupling $\gamma$, it is possible to explore different topological phase transitions within the several bulk gaps. These novel features  endow the topological phase diagram with an intrinsic richness and complexity, not present in other condensed-matter realizations of the QSH effect.

\begin{figure}[htb]
\begin{center}
\includegraphics[width=3.3in]{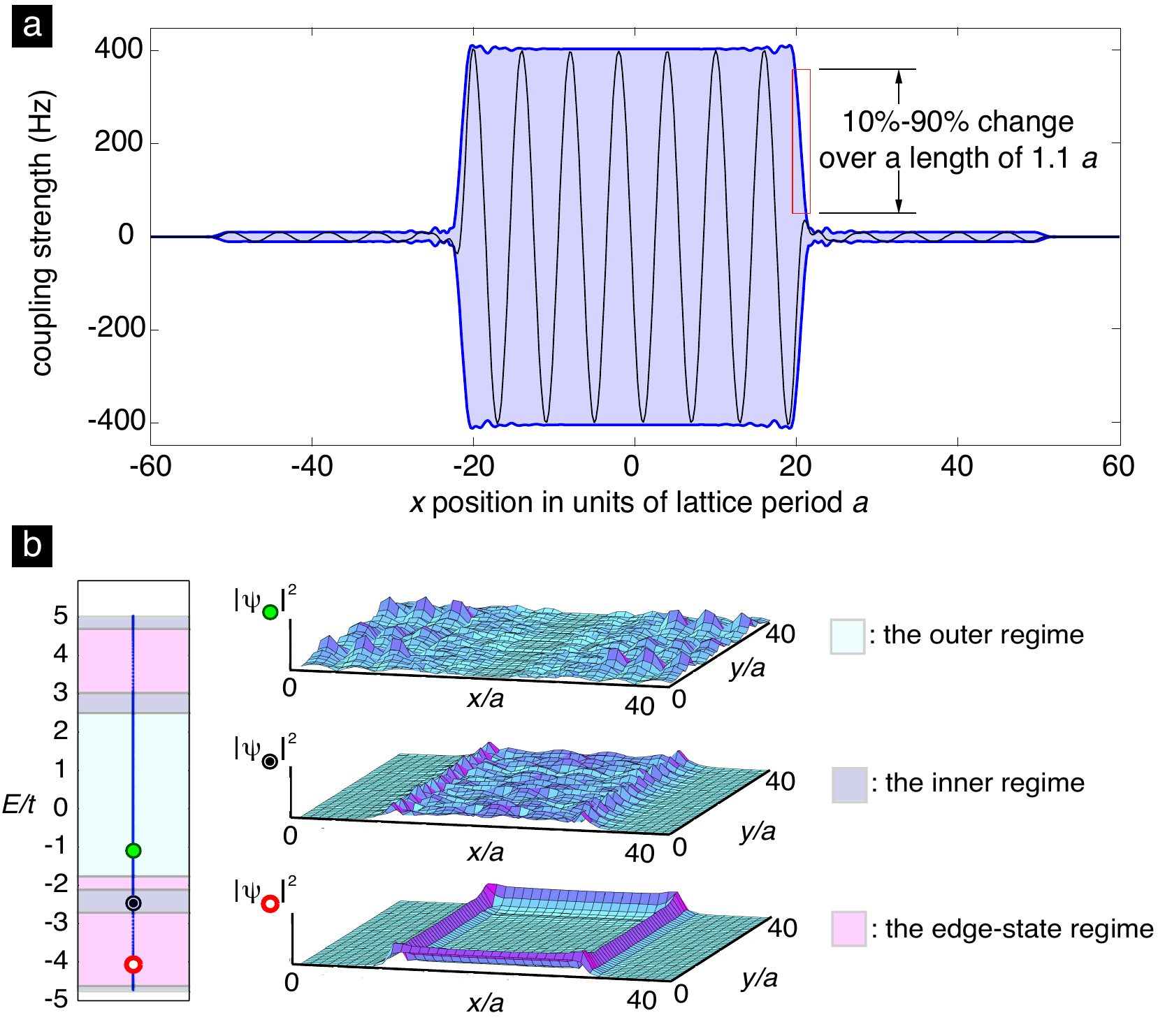}
\caption{\label{FIG4} (a) Position dependent coupling strength. The shaded blue region displays the position dependent coupling strength averaged over one period of oscillation of the RF fields.  The black trace depicts the Zeeman shift from the computed magnetic field along $\hat x$ which produces the RF coupling (at a representative time). (b) Discrete energy spectrum (blue dots) and typical amplitudes $\vert \psi_{\uparrow} (x,y)\vert^2$ in the presence of a harmonic potential and anisotropic hopping. The open lattice has $42 \times 42$ sites, $\gamma=\lambda_{\txt{stag}}=0$ and $\alpha\!=\!1/6$. The harmonic potential $V_{\txt{conf}}(x,y) \propto x^2 + y^2$ is such that $V_{\txt{conf}} (42,42)= 0.5 t$ and the hopping parameters of the inner and outer regions are respectively $t'_y=2 t$ and $t_y=0.1 t$. The amplitudes $\vert \psi_{\uparrow} (x,y)\vert^2 $ mark three distinct regimes in the associated spectrum: the outer (light green), the inner (dark blue) and the edge-state (magenta) regimes.}
\end{center}
\end{figure}

\vspace{-0.3cm}

We now describe a new feasible scheme to engineer a sharp interface where edge states can be localized. This is essential for detecting topological states in optical lattices, where the indispensable harmonic trap used to confine atoms destroys the edge states
when $V_{\txt{conf}} (\txt{edge}) \sim \Delta$, $\Delta$ being the bulk gap's width~\cite{Stanescuarxiv}. The key aspect of our proposal exploits the fact that the hopping along the $y$-direction, $t_y$, is controlled by spatially periodic RF transitions and hence can be tuned. Since the harmonic trap has a minimal effect at the center of the trap, we divide the chip into three regions: the central region is characterized by a large hopping $t'_y$, while the two surrounding regions feature small hoppings $t_y\ll t_y'$.  This can be realized by abruptly changing the current in the ``Raman'' wires to a much smaller value, thereby reducing the coupling matrix element on the single lattice site scale [Fig.~\ref{FIG4}(a)].  The resulting highly anisotropic hopping creates a sharp interface where the edge states of the central region -- a topological insulator -- localize. By controlling the strength of the Raman coupling, we squeeze the energy bands describing the outer parts so that they do not interfere with the bulk gaps of the central topological phase. We show that the helical edge states are robust and exactly localized at the designed \emph{interface}. Since the topological phases are confined in the center of the trap, one verifies that the phase diagrams discussed above are valid for a much wider range of the harmonic potential's strength. Fig.~\ref{FIG4}(b) illustrates the discrete energy spectrum and the typical wave-functions marking three distinct regimes. When the Fermi energy lies in the \emph{edge-state} regime,  the edge states are robustly localized within the interface. These edge states are composed of counter-propagating spin textures, in direct analog to the spin-polarized chiral edges states of integer QH systems.  That these edge states are chiral was unambiguously measured by detecting a ``pulse'' of electrons as they propagated around -- and not through -- the sample~\cite{Ashoori1992}.  It is simple to revisit this measurement in cold atom systems by using a focused laser beam to launch edge excitations, followed by spin-sensitive imaging. To confirm the topological insulating state, this excitation must travel along the edge with the two spin-components traveling in opposite directions. Further evidence of these edge-states could be probed through light Bragg scattering~\cite{Liu2010,Stanescuarxiv}. When $\gamma\!=\!0$, the spin Hall conductivity is given by $\sigma_s\!=\!e (N_{\uparrow}\txt{mod}2 ) /2 \pi$, where $N_{\uparrow}$ is a Chern number, which can be evaluated through the Streda formula applied to the spin-up density~\cite{Umucalilar2008}. The anisotropy $t'_y>t$ leads to bulk gaps of the order $\Delta \!\sim \! t$, requiring cold, though realistic, temperatures $T\!\sim \!10\nK$ to detect the QSH phase. 

We have described a concrete and realistic proposal of synthetic gauge fields in  optical lattices that overcomes the severe drawbacks affecting earlier schemes. We showed how such gauge fields are ideally suited to experimentally realize the most transparent QSH phase and hence allow to explore the validity of the $Z_2$-classification against interaction and disorder~\cite{Kane2005bis,edgebis}. In our multi-band system, a staggered potential is shown to drive gap-dependent QPT's which constitute a unique and rich feature. The cold-atom realization of topological band-insulators and helical metals proposed in this paper will pave the way for engineering correlated topological superfluids and insulators. Considering  atoms with more internal states, it is possible to  envisage situations where the edge states present a richer spin structure, and thus offer the opportunity to explore new avenues and exotic topological phases. 

N.G. thanks the F.R.S-F.N.R.S. M.L. thanks the Grants of Spanish MINCIN (FIS2008-00784 and QOIT), EU (NAMEQUAM), ERC (QUAGATUA) and of  Humboldt Foundation. A.~B. and M.-A. M.-D. thank the Spanish MICINN grant FIS2009-10061, CAM research consortium QUITEMAD, European FET-7 grant PICC, UCM-BS grant GICC-910758 and FPU MEC grant. I.S. and P.N. are supported by ONR grant N00014-09-1-1025A, grant 70NANB7H6138, Am 001 by NIST. I.B.S. is supported by the ARO with funds from the DARPA OLE Program, and the NSF through the PFC at JQI. 

\end{document}